\documentstyle[12pt]{article}

\documentstyle{titlepage}
\title{ON THE RELATIVISTIC QUANTUM FORCE}
\author{\bf ALI\ SHOJAI$^*$\ \&\ MEHDI\ GOLSHANI$^{**}$\\
Department of Physics, Sharif University of Technology\\P.O.Box 11365-9161 Tehran, IRAN\\
and\\Institute for Studies in Theoretical Physics and Mathematics,\\P.O.Box 19395-5531, Tehran, IRAN\\
$^*$Email: SHOJAI@PHYSICS.IPM.AC.IR\\
\ \ \ \ \ \ \ \ \ SHOJAI@NETWARE2.IPM.AC.IR\\
$^{**}$Fax: 98-21-8036317\\
\ \ \ \ \ \ \ \ \ 98-21-8036319}
\date{}
\begin{document}
\begin{bf}
\maketitle
\vspace{1cm}
\begin{center}
{\Large ON THE RELATIVISTIC QUANTUM FORCE}\\
{\bf A. Shojai \& M. Golshani}
\end{center}
\vspace{0.5cm}
\begin{center}
{\bf ABSTRACT}
\end{center}
{\it In the extension of the de-Broglie--Bohm causal quantum theory of motion 
to the relativistic particles, one faces with serious problems, like the problem
of superluminal motion. This forces many authors to believe that there is not
any satisfactory causal theory for particles of integer spin. In this paper, it is shown
that the quantal behaviour is the result of direct-particle-interaction of the particle
with all of its possibilities. The formulation is, then, extended to the relativistic
particles of arbitrary spin. The presented theory has the following advantages.
(1) It leads to a deeper understanding of the quantal behaviour. (2) It has
no superluminal motion. (3) It is applicable to any spin. (4) It provides a
framework for understanding the problem of creation and annihilation of particles. (5) It
provides a framework for understanding the spin--statistics relationship. (6) It
does not need the two fundamental assumptions of the de-Broglie--Bohm quantum theory of motion, i.e., the
guiding-formula postulate and the statistical postulate.}
\vspace{1.5cm}
\section{INTRODUCTION}
\par
In the beginings of this century, some experimental results were achieved which
{\it seemed\/} to be in contradiction with classical mechanics. An important
one was the problem of the stability of atoms, in which the classicaly expected motion
was not observed.
It is a well-known result that any atom consists of a nucleus with positive electric charge and
some negatively charged electrons distributed around it. According to classical physics,
one expects that the electrons circle around the nucleus and radiate untill they fall into it.
Thus, atoms should be unstable. In practice, it was seen that atoms are stable and do not
radiate as the theory predicts. In fact, there are situations in which atoms have not any
radiation at all. A great number of physicists departed from
logic
and jumped to the conclusion that, in the microscopic domain, it is
meaningless to speak of any path for particles. Copenhagenists had the 
following prescription: {\it Look at the system as a
black-box, then you can find some rules to relate output to input. No property
of the system has reality until it is observed. So only inputs and outputs 
are elements of reality and can be spoken about\/}.
\par
The Copenhagen formulation of quantum mechanics
has several departures from classical physics,
two of them are relavant to our discussion:
\par
(1)--- Bohr's complementarity principle, which states that a system may be 
represented by different quantities but with different descriptions.
Two different descriptions {\it must not\/}
be used simultaneously but any one can be used at a time, depending on the 
setup. In the Copenhagen formulation of quantum mechanics, 
this means that there are incompatible observables which their eigenspaces,
and any one can be chosen to represent the system. As a
result, incompatible observables satisfy Heisenberg's uncertainty principle,
so that they cannot be measured exactly at the same time.
Position and momentum are incompatible observables in the above sense,
and thus they cannot be measured simutaneously. Therefore it
is impossible to define path for particles in the Copenhagen quantum mechanics.
\par
(2)--- The measurment process cannot be explained in detail. Only it is stated
that, after the measurment of some observable, one of its eigenvalues is obtained
with a predictable probability distribution, after the system is being represented
by the corresponding eigenstate (the reduction of the state vector).
Why this is so? What is the extraordinary feature of the measurement device 
which enables it to reduce the state vector? If one considers the Copenhagen quantum
mechanics as a universal theory, one must use this theory to investigate the
behaviour of the measurement device. On doing this, one faces with the problem that
there is no reduction at all. Reduction may occure if the measurement device
is measured by another device. But what about this second measurment device?
In this way the measurement process is paradoxial. One way out of this, is to assume
that at some level the measurement device obeys classical physics and not the 
Copenhagen quantum mechanics. This assumption deprives the latter of a universal 
status. In addition, how could a theory which claims to be the underlying theory of
classical physics, depend on classical physics?
\par
Note that, we don't use the word
{\it Copenhagen interpretation\/} of quantum mechanics, because the so--called different interpretations of 
quantum mechanics are not merely different ways of interpreting something.
It is well-known that the Copenhagen quantum mechanics does not tell anything, say,
about the time of tunelling of a particle through some potential barrier -- it
is an unallowed question. In contrast, in the de-Broglie--Bohm quantum theory 
of motion, such questions are allowed and can be answered.(Holland 1993a)
\par
The Copenhagen quantum mechanics is a {\it black--box\/} theory. It takes
some initial conditions and gives final results. The only thing 
which has almost always a causal evolution is the state of the
system. During the measurment process the state of
the system has no causal evolution. The situation is just like the investigation of a circuit element in
electrical engineering. There, one calculates the scattering matrix of the circuit element,
using an effective version of Maxwell's equations, the so--called KVL and KCL 
laws. In this sense, the Copenhagen quantum mechanics is an effective theory of a much deeper theory.
It can be used to calculate the scattering matrix of the system. There must exist
an underlying physical theory, which leads to the Copenhagen quantum
mechanics as an effective black--box theory.
\par
The de-Broglie--Bohm quantum theory of motion (Bohm 1952a,b) has the general
plan to describe any system in a space--time background causally, i.e. it considers
path for particles and describes all physical processes, including measurment
process, event by event. This theory may be considered as the underlying theory 
of the Copenhagen quantum mechanics. If one restricts himself to the so--called
{\it observables\/} of the Copenhagen quantum mechanics (i.e. inputs and
outputs), this latter theory would emerge from the former, but in fact 
the de-Broglie--Bohm quantum theory of motion predicts more. It is not merely a 
black box theory, it gives {\it explanation\/} rather than {\it description\/}. It can be seen 
that the de-Broglie--Bohm quantum theory of motion has testable predictions 
beyond the Copenhagen quantum mechanics.(Holland 1993a)
\par
In this theory, a particle is always guided by the phase of an {\it objectively real\/}
field --- the wave-function. On the other hand, the norm of the wave-function determines 
the ensemble density of the particle.
Although, in the nonrelativistic limit, the de-Broglie--Bohm quantum theory of motion 
works successfully, its generalization to the relativistic domain has
problems. For the case of integer spin it seems that particle theory does not work,
because of tachyonic solutions. This has forced many physicists (Holland 1993b, Lam 1994a,b, Bohm 1993) to believe
that we have integer-spin fields and  half-integer-spin particles.
\par
In a recent work (Shojai 1996), we have shown that the quantum field (the wave-function), 
as any other field, may be viewed as direct-particle-interaction and thus there is no need to introduce the wave-function, as an
objectively real field. This 
point would be clarified in the next section. In other sections, we shall 
develop a relativistic quantum theory of motion based on the direct-particle-interactions.
The suggested theory, has no superluminal motion. It does not need the two 
fundamental assumptions of Bohm, the guidance formula (i.e. momentum is the gradiant
of the phase of the wave-function) and the statistical assumption (i.e. the
probability density is the norm of the wave-function). Our theory, as it will 
be shown, is applicable to any spin, and to curved space-time.\\
\section{THE \ QUANTUM \ FORCE \ AND \ DIRECT PARTICLE INTERACTIONS}
\par
In the de-Broglie--Bohm quantum theory of motion, the wave-function which is an objectively
real field, exerts a force (usually called the quantum force) on the particle,
producing the classically unexpected motions. Although, in introducing the 
de-broglie--Bohm quantum theory of motion, it is customary to begin with the 
wave-function and the Schr\"odinger equation, we do not follow this line of
approach. This is because we do not beleive in the wave-function as an objectively
real field and our aim is to derive the quantum force from direct-particle-interactions.
So we introduce here an approach which directly leads to the quantum force
and which seems to us more logical.
\par
Let us return to a question which was prevalent in the begining of this century, i.e., why atoms 
are stable? If one wants to remain faithfull to the logic, one must argue that:
{\it If atoms do not radiate, electrons have no accelerated motion. And because of
homogeneity and isotropy of space, if electrons have any motion, it must be a radial one
with constant velocity. But this leads to the evaporatin of atoms. So the conclusion is
that electrons are at rest in the stable atoms. Accordingly, one expects to have some
force which balances the coloumb force between electrons and nucleus, and
we call this the quantum force}. Here one discovers a new force, in the same way Newton
discovered gravity. In many other experiments the situation is completely similar to the above case.
\par
Now our task is to determine the law of the quantum force. We assume that it
can be obtained from a potential -- the quantum potential. We categorize
the results of experiments in three facts, and then use them to obtain the form of the 
quantum potential.\\
\par
{\bf FACT 1:} {\it For almost all classical potentials, we have seen cases where the particle
is \ \ \ \ \ \ \ \ \ \ \ \ \ completely at rest.\/}\\
This fact, as we saw in the foregoing example, indicates that there is a new potential
-- the quantum potential ($Q$). If the classical and quantum forces cancel each other,
the particle may be at rest at any position. This is an example of the so-called {\it steady state\/}. Clearly 
under these conditions the quantum potential is not a function of kinematical parameters like momentum etc. In 
addition,  the quantum potential can not be a pre-defined function of position, 
otherwise, it cannot cancel {\it almost any\/} classical potential. 
The only quantity on which the quantum potential can depend, is the position at which the particle is at rest. 
In laboratory, it is very difficult to measure this position, it is 
customary to repeat the experiment a large number of times and use the ensemble density of the particle at different positions.
So the quantum potential is a function of $\rho(\vec{x})$, the density of particles at rest
\begin{center}
FACT 1 $\Longrightarrow$ $Q$ depends on $\rho$
\end{center}
\par
{\bf FACT 2:} {\it Different locations are correlated through the quantum potential.\/}\\
It is an accepted matter that in the two-slit experiment, particles moving through the first slit, 
understand if the second is open or closed and vice versa. This means that the quantum potential is 
non-local, even for one particle systems. A sudden change in the boundary conditions,
acts at other places instantaneously. So the quantum potential must be related
to the derivatives of $\rho$. At this stage we assume that it depends only on
the first and second derivatives of $\rho$, later we shall relax this assumption.
\begin{center}
FACTS 1 and 2 $\Longrightarrow$ $Q$ depends on  $\rho$, $\vec{\nabla}\rho$ and $\nabla ^2 \rho$
\end{center}
\par
{\bf FACT 3:} {\it The total number of particles is irrelavant.\/}\\
If one performs an experiment with one hundered or with one thousand particles, the result is the same.
So the quantum potential will not change if one multiplies $\rho$ by a constant, i.e.,
it is a function of the {\it shape\/} of $\rho$.
\begin{center}
FACTS 1 and 2 and 3 $\Longrightarrow$ $Q$ depends on $\frac{\vec{\nabla}\rho}{\rho}$ and $\frac{\nabla ^2 \rho}{\rho}$
\end{center}
\par
Using these three facts and noting that the quantum potential must be rotationally invariant, one
concludes that in the steady state, i.e. when particles are at rest, the general form of the quantum potential is
\begin{equation}
Q=q\left ( \frac{\nabla ^2 \rho}{\rho} +a\frac{|\vec{\nabla}\rho|^2}{\rho^2} \right )
\end{equation}
Where $q$ and $a$ are two constants and our task is to determine their values. Note that for the hydrogen atom, on the basis
of symmetry considerations, one may suggest the distribution $\rho \sim e^{-\alpha r}$. Then the total energy $E=Q+V=Q-e^2/r$ is a constant provided
$\alpha=-\frac{e^2}{2q}$. This shows the possibility of stable atoms. Now consider a particle in the classical potential $V=\frac{1}{2} m\omega ^2 x^2$. Again one may 
suggest the density as $\rho \sim e^{-\alpha x^2}$, $\rho \sim x^2 e^{-\alpha x^2}$, and so on.
The first is possible if $4\alpha^2(1+a)q=-\frac{1}{2}m\omega ^2$ and the second is possible if $4\alpha^2(1+a)q=-\frac{1}{2}m\omega ^2$ and $a=-\frac{1}{2}$.
Experiments show that the difference between the energies of these two configurations is $\Delta E=\hbar \omega$, which leads to $q=-\frac{\hbar ^2}{4m}$.
Thus, we have
\begin{equation}
Q=\frac{-\hbar ^2}{4m}\left ( \frac{\nabla ^2 \rho}{\rho} -\frac{1}{2}\frac{|\vec{\nabla}\rho|^2}{\rho^2} \right )=\frac{-\hbar ^2}{2m}\frac{\nabla ^2 \sqrt{\rho}}{\sqrt{\rho}}
\end{equation}
Although we obtained  this result for the steady state, we now generalize it to any case. That is, we assume that {\it any particle 
either moving or at rest is always acted on via the quantum potential given by (2)}. Note that (2) is written
in terms of $\rho$, the density of an ensemble of the particle under consideration, so {\it the quantum potential represents the interaction
of a particle with all of its possibilities}.
\par
The reader familiar with the de-Broglie--Bohm quantum theory of motion remembers that
(2) is in fact Bohm's quantum potential (Bohm 1952a,b, Bohm 1993, Holland 1993a). In that theory, this potential is produced as an interaction between the particle 
and the {\it objectively real\/} wave-function field. In our view, however, the quantum 
potential as well as any other potential must be considered as the {\it direct-particle-interaction\/} and not 
as the {\it field-particle-interaction\/}. We have argued in favour of the first previousely(Shojai 1996), but here is another argument.
Consider for a moment the case of electromagnetism. One may considers the electromagnetic fields as produced by charges and currents and write integral relations
to obtain these fields from charges and currents. Then one uses these electromagnetic fields to calculate the Lorentz force exerted on a charged particle.
In this view the electromagnetic fields are merely {\it mathematical tools\/}, they are not objectively real fields. If one sets the charge and current densities equal to zero,
the electromagnetic fields would be zero(Ribari\v c 1990). Another way, is to give to the electromagnetic fields 
an objectively real character and to write for them a set of differential equations -- the so-called Maxwell's equations. 
A remarkable property of Maxwell's equations is that, on setting the charge and current densities equal to zero, the electromagnetic fields do not necessarily go to zero. It is possible to have electromagnetic 
fields without any sources. This is a result of the fact that they are objectively real in this view.
\par
Thus there is a fundamental difference between direct-particle-interaction and field-particle-interaction pictures. The latter gives some additional solutions which are not clear that if
they exist in nature or not. The situation is more important in the case of quantum force. In the de-Broglie--Bohm quantum theory of motion, the field equation is the Schr\"odinger equation which is sourceless,
but one puts by hand this assumption that the wave-function exists if and only if the particle exists. One should not confuse this case with the case in which the wave-function consists of two packets, one {\it empty\/} and one
accompanied by a particle. If there is no particle at all, there is no wave-function. 
Thus, in the case of the quantum force, sourceless fields do not exist, and so direct-particle-interaction picture is more suitable.
According to this reasoning and some others (Shojai 1996), we beleive that the quantum force is a result of direct-particle-interactions and not a result of a field. This is our departure from the de-Broglie--Bohm
quantum theory of motion, which as we shall see, enables us to avoid two fundamental assumptions of Bohm's theory, i.e. the guiding formula and the statistical postulate. These are now consequences of the equations of motion.
Furthermore, this view enables us to write a succesful causal theory for the relativistic quantum systems.
\par
Let us write, as a first step towards the final goal, an effective lagrangian theory for the nonrelativistic quantum force. By effective we mean that the quantum force is introduced by hand in the form (2) in the lagrangian.
Later we derive this from direct-particle-interactions. Consider an ensemble of similar particles with density $\rho(\vec{x},t)$ and the Hamilton-Jacobi function ${\cal S}(\vec{x},t)$. We use the Hamilton-Jacobi formalism, because it is more suitable
for describing an ensemble of particles. The action is(Holland 1993a)
\begin{equation}
{\cal A}=\int dt \left [ \rho \left ( \frac{\partial {\cal S}}{\partial t}+\frac{|\vec{\nabla}{\cal S}|^2}{2m}+V+Q \right ) \right ]
\end{equation}
where $V$ is the classical potential.  Variation of this action leads to the following equations of motion
\begin{equation}
\frac{\partial {\cal S}}{\partial t}+\frac{|\vec{\nabla}{\cal S}|^2}{2m}+V+Q=0
\end{equation}
\begin{equation}
\frac{\partial \rho}{\partial t}+\vec{\nabla}\cdot \left ( \rho \frac{\vec{\nabla}{\cal S}}{m} \right )=0
\end{equation}
The relation with the Copenhagen quantum mechanics can be understood by the following canonical transformation
\begin{equation}
(\rho,{\cal S})\longrightarrow (\psi, \psi ^*);\ \ \ \ \ \ \ \ \psi=\sqrt{\rho}\ e^{i{\cal S}/\hbar}
\end{equation}
In this way we obtain the Schr\"odinger equation instead of (4) and (5).
The relativistic extension of  this formalism is straightforward, noting that time and space coordinates must
be treated on the same foot, and that in the nonrelativistic case the kinetic energy 
$|\vec{\nabla}{\cal S}|^2/2m$ is modified by $Q$. The result is
\begin{equation}
{\cal A}=\int d^4x \left [ \frac{1}{2} \rho \left ( \partial _{\mu}{\cal S} \partial ^{\mu}{\cal S}+Q \right ) \right ]
\end{equation}
The equations of motion are now
\[ \partial _{\mu}{\cal S} \partial ^{\mu}{\cal S}=-Q;\ \ \ \ \ or\] 
\begin{equation} 
\frac{dP_{\mu}}{d\tau}=\frac{-1}{2{\cal M}}\partial _{\mu}Q\ \ \ \ \ where\ \ P_{\mu}=-\partial _{\mu}{\cal S}
\end{equation}
\begin{equation}
\partial _{\mu}(\rho \partial ^{\mu}{\cal S})=0
\end{equation}
On multiplying (8) by $P^{\mu}$ one sees that we must interpret the quantum potential as it is proportional to the mass squared of the particle, i.e.
\begin{equation}
Q=-{\cal M}^2c^2
\end{equation}
The relativistic extension of (2) is
\begin{equation}
-Q={\cal M}^2c^2=m^2c^2+\hbar^2\frac{\Box\sqrt{\rho}}{\sqrt{\rho}}
\end{equation}
where we have introduced the constant $m^2c^2$ as the classical limit of ${\cal M}^2c^2$.
Therefore the mass is variable and is the source of the quantum force. It is worthwhile to note that this variable
mass has nothing to do with that variable mass expressed sometimes in relativity. ${\cal M}$ is the rest mass, and its variation produces the quantum force.
\par
The action (7) has the problem that we must use (11) as well as the equations of motion, i.e. by simply varying (7) all the equations are not derived. To overcome this difficulty, 
we introduce an {\it auxiliary field\/} $\Lambda (x)$ and write the effective action as
\begin{equation}
{\cal A}=\int d^4x \left [ \frac{1}{2} \rho \left ( \partial _{\mu}{\cal S} \partial ^{\mu}{\cal S}-{\cal M}^2c^2 \right )+\frac{1}{2}\partial _{\mu}\Lambda \partial ^{\mu}\Lambda -\frac{1}{2}\frac{c^2}{\hbar^2}(m^2-{\cal M}^2)\Lambda ^2 \right ]
\end{equation} 
The variation of this leads to 
\begin{equation}
\partial _{\mu}{\cal S} \partial ^{\mu}{\cal S}={\cal M}^2c^2
\end{equation}
\begin{equation}
\partial _{\mu}(\rho \partial ^{\mu}{\cal S})=0
\end{equation}
\begin{equation}
\rho=\frac{\Lambda^2}{\hbar^2}
\end{equation}
\begin{equation}
{\cal M}^2=m^2+\frac{\hbar ^2}{c^2}\frac{\Box\Lambda}{\Lambda}
\end{equation}
If we eliminate $\Lambda$ in terms of $\rho$ by (15), we get equations (8), (9) and (11). It is worthwhile to 
note that the relation with the relativistic Copenhagen quantum mechanics can be seen via the canonical transformation
\begin{equation}
(\Lambda, {\cal S})\longrightarrow (\phi, \phi ^*);\ \ \ \ \ \ \ \ \phi=\Lambda \ e^{i{\cal S}/\hbar}
\end{equation}
The result is the Klein-Gordon equation instead of (13)-(16). Note that, since $\Lambda$ is an auxiliary field,
the wave-function $\phi$ is not an objectively real field, in contrast to the de-Broglie--Bohm quantum theory of motion.
\par
In the Appendix A, we have shown that the nonrelativistic quantum potential can be derived from a specific 
direct-particle-interaction. It consists of three parts. The first part reflects the basic
property of any direct-particle-interaction, i.e. its {\it very\/} dependence on the location of the particles.
Thus the first part is an expression, that ensures that each particle is at its right position, i.e. the position derived from the equations of motion.
In (Shojai 1996) we have shown that such a nonordinary term may be written as $1/\sum_{i=1}^{N}\delta(\vec{x}-\vec{a}_i(t))=1/\rho(\vec{x},t)$, where $\vec{a}_i(t)$ is the right position of the $i$th particle
of an ensemble of similar particles. 
The second and third parts are two exponential interactions, one short range and the other long range. So the direct-particle-interaction is {\it chosen} to be
\begin{equation}
Q(\vec{x},t)=\frac{U_0}{\sum_{i=1}^{N}\delta (\vec{x}-\vec{a}_i(t))}
\left [ \sum_{j=1}^{N}\exp \left \{ -\frac{|\vec{x}-\vec{a}_j(t)|^2}{\alpha _s^2} \right \} \right ]
\left [ \sum_{k=1}^{N}\exp \left \{ -\frac{|\vec{x}-\vec{a}_k(t)|^2}{\alpha _{\ell}^2} \right \} \right ]
\end{equation}
In this relation $N$ is the total number of the particles, $U_0$ is some constant, $\alpha _s$ and $\alpha _{\ell}$ are the ranges of
short and long range interactions respectively. It may seem that this is not a well-defined function
because of the appearance of delta functions in the denominator. But it must be noted that in practice, we replace
it with the smooth function $\rho(\vec{x},t)$.
It can be easily shown(Shojai 1996) (see also Appendix A) that the potential (18) can be written as
\begin{equation}
Q(\vec{x},t)=\frac{\Omega_0}{\rho (\vec{x},t)} \int d^3y\ \sqrt{\rho(\vec{x},t)}\ {\cal G}(\vec{x}-\vec{y})\ \sqrt{\rho(\vec{y},t)}
\end{equation}
where the kernel is given by
\begin{equation}
{\cal G}(\vec{x}-\vec{y})=\exp [-\beta^2|\vec{x}-\vec{y}|^2]
\end{equation}
and $\Omega_0$ and $\beta$ are constants related to $U_0$, $\alpha _s$ and $\alpha _{\ell}$. As it is shown in (Shojai 1996) and Appendix A, the integral (19) with the kernel (20) can be carried out leading to 
(provided $N$ is very large) (see the note at the end of Appendix A.1)
\[ Q(\vec{x},t)=\Omega_0 (\pi/\beta^2)^{3/2} \frac{1}{\sqrt{\rho (\vec{x},t)}} \left \{ \exp [\nabla ^2/4\beta ^2] \right \} \sqrt{\rho(\vec{x},t)} \]
\begin{equation}
=\Omega_0 (\pi/\beta^2)^{3/2} \left ( 1+\frac{1}{4\beta ^2}\frac{1}{\sqrt{\rho (\vec{x},t)}}\nabla^2\sqrt{\rho (\vec{x},t)}+ \cdots \right )
\end{equation}
which is the de-Broglie--Bohm quantum potential corrected by small terms. 
The importance and implications of these corrections to the de-Broglie--Bohm quantum
potential are discussed elsewhere (Shojai 1996).
The extension of this formalism, to the relativistic case will be done in the next section.
But before doing so, it is instructive to note that in order to fit (21) with (2) one must assume that $\Omega _0 <0$ so that $U_0 <0$. This means that our potential is $-\infty$ if the particle is not at its own location and is zero otherwise. That is

 say, 
the correct position of the particle represents an absolute maximum of the potential. Any particle
at $\vec{x}\neq \vec{a}_i(t)$ and with finite energy will move with infinite velocity towards $\vec{x}=\vec{a}_i(t)$ and on reaching it, it will move with velocity $\dot{\vec{a}}_i(t)$, otherwise the particle will go to infinity and
we shall not see any particle. Thus, although $\vec{x}=\vec{a}_i(t)$ is a maximum of the potential, the particle must be either at infinity (in which there is no observable particle) or at $\vec{x}=\vec{a}_i(t)$ with velocity $\dot{\vec{a}}_i(t)$.
\par
Another possibility yet exists. It is logical to assume that $U_0>0$ or $\Omega_0>0$ is also an allowed choice.
In this case, the potential is $+\infty$ if $\vec{x}\neq \vec{a}_i(t)$ and is zero if $\vec{x}=\vec{a}_i(t)$. Now a particle with finite energy, cannot be
in the region $\vec{x}\neq \vec{a}_i(t)$. It will move with velocity $\dot{\vec{a}}_i(t)$ at $\vec{x}=\vec{a}_i(t)$. 
Summing up, there are two cases both leading to the fact that particle will move with velocity $\dot{\vec{a}}_i(t)$ at $\vec{x}=\vec{a}_i(t)$.
Either $\Omega_0<0$ which we call the particle in this case a {\it particle\/} or 
$\Omega _0>0$ in which case we call the particle an {\it antiparticle\/}. This terminology
would be soon clarified. The difference between particle and antiparticle is in the sign of the quantum potential. So we have
\begin{equation}
\begin{array}{lr}
m\frac{d^2\vec{x}}{dt^2}=-(+1)\vec{\nabla}Q\ \ \ \ \ &for\ \ particles,\\
m\frac{d^2\vec{x}}{dt^2}=-(-1)\vec{\nabla}Q\ \ \ \ \ &for\ \ antiparticles.
\end{array}
\end{equation}
when $Q$ is given by (2). One can absorb this minus sign in the definition of momentum and assume that
\begin{equation}
\begin{array}{lr}
\vec{p}=+m\vec{v}\ \ \ \ \ \ \ &for\ \ particles,\\
\vec{p}=-m\vec{v}\ \ \ \ \ \ \ &for\ \ antiparticles.
\end{array}
\end{equation}
In the following section we shall extend this formalism to the relativistic domain.\\
\section{RELATIVISTIC QUANTUM FORCE}
\subsection{Spinless Particles}
\par
Now, on the basis of the discussions at the end of the previous section, it is a simple task to generalize (18) and (19)
to the relativistic case. The obvious generalization of (18) is as follows
\begin{equation}
Q(x)=\frac{U_0}{\sum_{i=1}^{N}\delta (x-a_i(\tau _i))}
\left [ \sum_{j=1}^{N}\exp \left \{ \eta _j \frac{(x-a_j(\tau _j))^2}{\alpha _s^2} \right \} \right ]
\left [ \sum_{k=1}^{N}\exp \left \{ \eta _k \frac{(x-a_k(\tau _k))^2}{\alpha _{\ell}^2} \right \} \right ]
\end{equation}
where $\tau _i$ represents the proper time of $i$th particle and $\eta _i$ is equal to $+1$
if $(x-a_i)^2$ is timelike and $-1$ otherwise (The signature of the metric
is chosen to be $+---$). Dealing with this form is very difficult, so we generalize 
relation (19) instead of (18) as follows
\begin{equation}
Q(x)=\frac{\Omega _0}{\rho (x)}\int d^4y\ \sqrt{\rho (x)}\ {\cal G}(x-y)\ \sqrt{\rho (y)}
\end{equation}
where
\begin{equation}
{\cal G}(x-y)=\exp [\beta ^2 (x-y)^2]
\end{equation}
The integral in (25) can be carried out leading to (see Appendix A)
\begin{equation}
Q(x)=\Omega_0 {\cal I} \frac{1}{\sqrt{\rho (x)}} \exp[-\Box/4\beta ^2] \sqrt{\rho(x)} =\Omega_0 {\cal I}\left ( 1-\frac{1}{4\beta ^2}\frac{1}{\sqrt{\rho(x)}}\Box\sqrt{\rho(x)}+ \cdots \right )
\end{equation}
which is the quantum potential given by (11) corrected by small terms. 
${\cal I}$ is defined in Appendix A. Again as
before, if $\Omega _0<0$ we have particles and if $\Omega _0>0$, we have antiparticles. Since in the case
of particles $Q=-{\cal M}^2c^2$ and for antiparticles $Q=+{\cal M}^2c^2$ and because of (25) one sees that
${\cal M}^2$ is semi-positive definite.
{\it So the important conclusion is that our theory does not have the problem
of superluminal motion}.
\par
Let us see what the effective lagrangian is now. Since the quantum
potential has derivatives of any degree, the lagrangian must be so. The
corresponding action is
\begin{equation}
{\cal A} =\int d^4x \left \{ \frac{1}{2}\rho [\partial _{\mu}{\cal S}\partial ^{\mu}{\cal S}-{\cal M}^2c^2]-\frac{1}{2}\Omega _0 {\cal I}
\Lambda e^{-\Box/4\beta^2}\Lambda -\frac{1}{2}\frac{c^2}{\hbar^2}(m^2-{\cal M}^2)\Lambda ^2 \right \}
\end{equation}
The equations of motion are (13), (14), (15) and (see Appendix B)
\begin{equation}
{\cal M}^2=m^2+\frac{\hbar ^2 \Omega _0 {\cal I}}{c^2} \frac{e^{-\Box/4\beta^2}\Lambda}{\Lambda}
\end{equation}
\par
Two points must be noted. First, the quantum potential derived from the above
direct-particle-interaction, is directly related to $\rho$, not through the auxiliary field $\Lambda$.
Second, when one tries an action like (28) for the system, the statistical postulate of Bohm
would emerge. That is to say, direct-particle-interaction picture enables
one to avoid two essential postulates of Bohm -- the guidance formula and the 
statistical postulate. This is because in the direct-particle-interaction
picture, the wave-function has no role, it is a mathematical object. {\it It can be either
introduced or not\/}.
\subsection{Spin One-half Particles}
\par
As a second step, we generalize the foregoing formalism of spinless particles to the case of 
spin one-half ones. In this case one deals with four densities $\rho ^{particle}_{spin\ up}$, $\rho ^{particle}_{spin\ down}$,
$\rho ^{antiparticle}_{spin\ up}$ and $\rho ^{antiparticle}_{spin\ down}$. Three things must be noted. First, in writting a potential
like (25), it is more convenient to deal with these four densities at the same time. Second, in making square roots we are left free for a phase
factor. Finally, one must notice that the sign of $\Omega _0$ for particle and antiparticle is opposite. Therefore we introduce
\begin{equation}
\Gamma =\left ( \begin{array}{l} \sqrt{\rho ^{particle}_{spin\ up}}\\
                                 \sqrt{\rho ^{particle}_{spin\ down}}\\
                                 \sqrt{\rho ^{antiparticle}_{spin\ up}}\\
                                 \sqrt{\rho ^{antiparticle}_{spin\ down}}\\
\end{array} \right )
\end{equation}
and
\begin{equation}
\gamma _0 =\left ( \begin{array}{cccc} 1&0&0&0\\0&1&0&0\\0&0&-1&0\\0&0&0&-1\\
\end{array} \right )
\end{equation}
Accordingly, the natural generalization of (25) to the case of spin one-half is
\begin{equation}
Q(x)=\frac{1}{2}\frac{\Omega _0}{|\overline{\Gamma}(x)\Gamma (x)|}\left [ \int d^4y\ \overline{\Gamma}(x)\ {\cal G}(x-y)\ \Gamma(y) +C.C. \right ]
\end{equation}
where
\begin{equation}
\overline{\Gamma}=\Gamma ^{\dag} \gamma _0
\end{equation}
It may seem that the kernel is now
\begin{equation}
{\cal G}(z)=\exp[\beta ^2 (z_{\mu}1)( z^{\mu}1)]
\end{equation}
but in general, the kernel may mix different elements of $\Gamma$, so we set
\begin{equation}
{\cal G}(z)=\exp[\beta ^2 (z_{\mu}1+\epsilon _{\mu})( z^{\mu}1+\epsilon ^{\mu})]
\end{equation}
instead of (34), where $\epsilon _{\mu}$ are four constant $4\times 4$ matrices.
Therefore, we assume that the correct quantum potential for spin one-half particles is given by (32) and (34).
Evaluation of the integral in (32) is straightforward.
Using the results of Appendix A, one obtains
\begin{equation}
Q(x)=\frac{1}{2}\frac{\Omega_0 {\cal I}}{|\overline{\Gamma}\Gamma|}\left \{ \overline{\Gamma} \left [ e^{-\Box/4\beta ^2}+e^{\epsilon _{\mu} \partial ^{\mu}}-1 \right ] \Gamma + C.C. \right \}
\end{equation}
In order to have definite transformation laws, one must assume that $\Gamma$ transforms
like a Dirac 4-spinor, therefore we set
\begin{equation}
\epsilon _{\mu} = \varepsilon \gamma _{\mu}
\end{equation}
where $\varepsilon$ is a constant and $\gamma _{\mu}$ are Dirac matrices.
\par
The action principle is
\[ {\cal A} =\int d^4x \left ( \frac{1}{2}\rho [\partial _{\mu}{\cal S}\partial ^{\mu}{\cal S}-{\cal M}^2c^2] \right . \]
\begin{equation}
\left . -\frac{1}{4}\Omega _0 {\cal I} \left \{ \overline{\Gamma} \left [ e^{-\Box/4\beta ^2}+e^{\varepsilon \gamma _{\mu} \partial ^{\mu}}-1 \right ] \Gamma + C.C. \right \} -\frac{1}{2}\frac{c^2}{\hbar^2}(m^2-{\cal M}^2)|\overline{\Gamma}\Gamma|
\right )
\end{equation}
The equations of motion are (13), (14) and
\begin{equation}
\rho=\frac{|\overline{\Gamma}\Gamma|}{\hbar ^2}
\end{equation}
\begin{equation}
{\cal M}^2=m^2+\frac{\hbar ^2 \Omega _0 {\cal I}}{c^2}\frac{1}{|\overline{\Gamma}\Gamma|}\overline{\Gamma}\left [ e^{-\Box/4\beta ^2}+e^{\varepsilon \gamma _{\mu} \partial ^{\mu}}-1 \right ] \Gamma
\end{equation}
It is worhtwhile to notice that the last equation can be written as
\begin{equation}
\varepsilon \gamma _{\mu} \partial ^{\mu} \Gamma + \Gamma + corrections =0
\end{equation}
where similarity to Dirac's equation is apparent.\\
\subsection{Spin One Particles}
\par
Now that we have formulated the theory for spin one-half particles, the formulation
for spin one is obvious. Suppose that, for $\sqrt{\rho}$, we choose the four-vector $A_{\mu}$
with an auxiliary relation like
\begin{equation}
\partial _{\mu}A^{\mu}=0
\end{equation}
this provides a spin one representation of Lorentz group.
\par
The quantum potential may be written as
\begin{equation}
Q(x)=\frac{\Omega _0}{|A_{\mu}(x)A^{\mu}(x)|} \int d^4y\ A_{\nu}(x)\ {\cal G}^{\nu \kappa}(x-y)\ A_{\kappa}(y) 
\end{equation}
$\varepsilon$ terms like those in (34) cannaot exist, because there is no such matrices in this
case leaving  $Q$ invariant. So
\begin{equation}
{\cal G}^{\nu \kappa}(z)=g^{\nu \kappa}  \exp(\beta^2 z^2) 
\end{equation}
where $g^{\nu \kappa}$ is the metric of space-time. Using the results of Appendix A
we have
\begin{equation}
Q(x)=\frac{\Omega _0 {\cal I}}{|A_{\mu}A^{\mu}|} A_{\nu}e^{-\Box /4\beta^2} A^{\nu}
\end{equation}
The action principle is
\begin{equation}
{\cal A} =\int d^4x \left ( \frac{1}{2}\rho [\partial _{\mu}{\cal S}\partial ^{\mu}{\cal S}-{\cal M}^2c^2] -\frac{1}{2}\Omega _0 {\cal I} A_{\mu}e^{-\Box/4\beta ^2} A^{\mu} -\frac{1}{2}\frac{c^2}{\hbar^2}(m^2-{\cal M}^2) |A_{\mu}A^{\mu}| \right )
\end{equation}
The equations of motion are (13), (14) and
\begin{equation}
\rho=\frac{|A_{\mu}A^{\mu}|}{\hbar ^2}
\end{equation}
\begin{equation}
{\cal M}^2=m^2+\frac{\hbar ^2 \Omega _0 {\cal I}}{c^2}\frac{1}{|A_{\mu}A^{\mu}|}A_{\nu} e^{-\Box/4\beta ^2}A^{\nu}
\end{equation}
\par
We end this section by summarizing its results. As we saw, the quantum potential
can be derived from a direct-particle-interaction of the form
\begin{equation}
Q(x)=\frac{\Omega _0}{\rho (x)} \int d^4y\ \sqrt{\rho (x)}\ {\cal G}(x-y)\ \sqrt{\rho (y)} 
\end{equation}
where
\begin{equation}
\sqrt{\rho}=\left \{ \begin{array}{ll} \Lambda & spin = 0\\
                                       \Gamma  & spin = \frac{1}{2}\\
                                       A_{\mu} & spin = 1\\
                      \end{array} \right .
\end{equation}
and ${\cal G}$ is an appropriate kernel. If one writes an effective lagrangian 
in terms of the auxiliary fields $\Lambda$, $\Gamma$ or $A_{\mu}$, the equations
of motion are
\begin{equation}
\frac{d P_{\mu}}{d \tau}=-\frac{1}{2 {\cal M}}\partial _{\mu}Q
\end{equation}
\begin{equation}
\partial _{\mu}(\rho P^{\mu})=0
\end{equation}
\begin{equation}
\rho = \frac{1}{\hbar ^2}\left \{ \begin{array}{ll} \Lambda ^2 & spin=0\\
|\overline{\Gamma} \Gamma| & spin=\frac{1}{2}\\ |A_{\mu}A^{\mu}| & spin=1\\ \end{array} \right .
\end{equation}
\begin{equation}
{\cal M}^2 = m^2+ \frac{\hbar ^2 \Omega _0 {\cal I}}{c^2}\frac{1}{\rho}\left \{ \begin{array}{ll} \Lambda e^{-\Box/4\beta^2}\Lambda & spin=0\\
\overline{\Gamma}\left [ e^{-\Box/4\beta^2}+e^{\varepsilon \gamma _{\mu} \partial ^{\mu}}-1 \right ] \Gamma  & spin=\frac{1}{2}\\
A_{\mu}e^{-\Box/4\beta^2} A^{\mu} & spin=1\\ \end{array} \right .
\end{equation}
The last equation, upon expansion, can be written as
\begin{equation}
\left \{ \begin{array}{ll} Klein-Gordon\ equation\ + \ corrections, \ \ \ \ & spin=0\\
Dirac\ equation\ + \ corrections, \ \ \ \ & spin=\frac{1}{2}\\
Maxwell\ equations\ + \ corrections, \ \ \ \ & spin=1\\ \end{array} \right .
\end{equation}
\section{OBSERVATIONS}
\par
In the previous sections, we have obtained that quantal behaviours are the
results of a new kind of force between different particles in an ensemble, or in other words,
the direct interaction of a particle with all of its possibilities. The
specific form of this direct particle interaction is derived both for the
nonrelativistic and the relativistic domains. Here we have some observations:
\begin{enumerate}
\item It is instructive to write the conseved currents in terms of the auxiliary
fields defined in the previous section. In doing so, one must be careful. Since our
lagrangians contain derivatives of any order, the standard version of Ne\"other's
theorem is not applicable. In Appeendix B, we have derived a generalized version
of Ne\"other's theorem which is useful for such lagrangians.
\item The second point which is more important, is about the mass conservation
equation (52). As its name indicates, it is a relation for conservation of mass
\begin{equation}
\partial _{\mu}(\rho {\cal M} U^{\mu})=0\ \ \ \ \ or\ \ \ \ \ \frac{d}{dt}\int d^3x\ \rho {\cal M} U^0 =0
\end{equation}
where $U_{\mu}$ is the four-velocity.
\par
For charged particles, the electric current must also be conserved. Thus,
the electric charge (${\cal E}$) is proportional to ${\cal M}$. That is
\begin{equation}
{\cal C}_{\mu}=\rho {\cal E} U_{\mu} ;\ \ \ \ \ \ \partial_{\mu}{\cal C}^{\mu}=0 ;\ \ \ \ \ \ \frac{{\cal E}}{{\cal M}}=\frac{e}{m}
\end{equation}
\par
The number density of particles defined as
\begin{equation}
{\cal N}_{\mu}=\rho U^{\mu}
\end{equation}
is not conserved, because using (56) we have
\begin{equation}
\partial _{\mu}{\cal N}^{\mu}=-\frac{1}{{\cal M}}{\cal N}^{\mu}\partial _{\mu}{\cal M}
\end{equation}
Therefore, one arrives at the important conclusion that {\it in the relativistic
quantum theory of motion, the number of particles and antiparticles is not
conserved. They can be created or annihilated\/}.
\par
In order to show how the creation and annihilation would emerge from
this theory, we consider the case of a charged spinless particle. For simplicity,
we only take into account the first and second terms of (27) and forget the corrections
to Maxwell equations. Also, we suppose that the electromagnetic interaction
is introduced as it is usuall, i.e. via minimal coupling and by introducing source
terms in in the Maxwell equations.
Accordingly, the governing equations are
\begin{equation}
(\partial _{\mu}{\cal S}-\frac{e}{c} A_{\mu})(\partial ^{\mu}{\cal S}-\frac{e}{c} A^{\mu})={\cal M}^2c^2
\end{equation}
\begin{equation}
\partial _{\mu}\left (\rho \left [\partial ^{\mu}{\cal S}-\frac{e}{c} A^{\mu} \right ] \right )=0
\end{equation}
\begin{equation}
\rho=\frac{\Lambda ^2}{\hbar ^2}
\end{equation}
\begin{equation}
{\cal M}^2=m^2+\frac{\hbar ^2}{c^2}\frac{\Box \Lambda}{\Lambda}
\end{equation}
\begin{equation}
\Box A_{\mu}=\frac{4 \pi}{c}\frac{e}{m}(\partial _{\mu}{\cal S}-\frac{e}{c} A_{\mu})
\end{equation}
As the initial conditions, we suppose that we have two packets of
the electromagnetic field moving twoards each other on the y-axis, i.e. we assume
\begin{equation}
A_{\mu}=(0,A,0,0)
\end{equation}
where
\begin{equation}
A=f(t) (A^++A^-)
\end{equation}
\begin{equation}
A^{\pm}=A(\xi ^{\pm})\ \ \ \ \ \ \ with\ \ \ \xi ^{\pm}=y\pm ct
\end{equation}
and $f$ is a function of time with the properties
\begin{equation}
f(t=-\infty)=1
\end{equation}
\begin{equation}
f(t=+\infty)=0
\end{equation}
The solution for $\Lambda$ is as follows
\begin{equation}
\Lambda=g(t) (\Lambda^++\Lambda^-)
\end{equation}
where
\begin{equation}
\Lambda^{\pm}=\Lambda(\xi ^{\pm})
\end{equation}
and
\begin{equation}
g(t=-\infty)=0
\end{equation}
\begin{equation}
g(t=+\infty)=1
\end{equation}
and that
\begin{equation}
\partial ^{\pm} \Lambda ^{\pm}=\frac{1}{\sqrt{2}}\frac{e}{c}A^{\pm}
\end{equation}
$\partial ^{\pm}$ represents differentiation with respect to $\xi ^{\pm}$.
Using the continuity equation (61) and the fact that the four-momentum is
defined as $P_{\mu}=-\partial _{\mu} {\cal S}+(e/c)A_{\mu}$, one obtains
\begin{equation}
P^0 \sim e^{-2\Lambda ^+} -e^{-2\Lambda ^-}
\end{equation}
\begin{equation}
P^2 \sim e^{-2\Lambda ^+} +e^{-2\Lambda ^-}
\end{equation}
Finally the Maxwell equations (64) lead to
\begin{equation}
P^1 =\frac{mc}{\pi e}(A^++A^-)(\partial ^+\partial ^-f)\  e^{-2(\Lambda ^++\Lambda ^-)g}
\end{equation}
which has the property
\begin{equation}
P^1(t=-\infty) =P^1(t=+\infty) =0
\end{equation}
\par
The picture one draws from the above peculiar solution is as follows
\subitem (a)-- At $t=-\infty$, there are only two electromagnetic packets, i.e.
two groups of photons moving towards each other.
\subitem (b)-- In the intermediate times, particles are created moving
longitudally and transversally. They produce electromagnetic fields and
reduce the initial electromagnetic fields to zero. (In other words, photons
are annihilated.) Then, they fly far away.
\subitem (c)-- At $t=+\infty$, there are only two packets of charged particles
flying in different directions, one with positive and one with negative energy. (Note
the minus sign in (75).) So particles and antiparticles are created.
\item In the previous section we saw that the present theory is mathematicaly
equivalent to the standard relativistic wave equations plus some corrections. 
And in the above, it was shown that this theory also presents a framework
for understanding the problem of creation and annihilation of particles. Therfore
it can be seen that the present theory is equivalent to the standard quantum field theories,
up to some corrections, if one restricts himself to the {\it observed quantities\/} of the Copenhagen
quantum mechanics. But it is worthwhile to note that it is by no means  another interpretation of quantum field theories.
\item One must note that all of the above formulation is applicable to the case of curved
space-time. One must simply replace the Minkowski metric with the
general metric $g_{\mu \nu}$ and $\partial _{\mu}$ with the covariant derivative
$\nabla _{\mu}$, in order to account for the interaction with gravity. The metric
$g_{\mu \nu}$, in a semi-classical approach would be
determined by the Einstein equation
\begin{equation}
{\cal R}_{\mu \nu}-\frac{1}{2}{\cal R} g_{\mu \nu}=\kappa {\cal T}_{\mu \nu}
\end{equation}
where ${\cal T}_{\mu \nu}$, the energy-momentum tensor, is derived in the Appendix B.
At a more deeper level, one must deal with $g_{\mu \nu}$ as other auxiliary fields.
That is to say, gravitons are represented by the tensor
representation for $\sqrt{\rho}$, i.e. $\sqrt{\rho}\sim h_{\mu \nu}=g_{\mu \nu}-\eta _{\mu \nu}$.
Note that if one wants to look at gravity in a non-geometric manner, it is
described by $h_{\mu \nu}$ not $g_{\mu \nu}$. (See e.g. (Ohanian 1976))
As in the previous section, the direct-particle-interaction leading to the
mass function ${\cal M}$ of gravitons is given by
\begin{equation}
Q(x)=\frac{\Omega _0}{|h_{\mu \nu}(x)h^{\mu \nu}(x)|}\int d^4y\ h_{\alpha \beta}(x)\ {\cal G}^{\alpha \beta \gamma \delta}(x-y)\ h_{\gamma \delta}(y)
\end{equation}
where ${\cal G}^{\alpha \beta \gamma \delta}(x-y)$ is an appropriate kernel.
Note that in the above relation we must use $\eta _{\mu \nu}$ for raising and
lowering the indices.
\item It is worthwhile to note that high degree derivatives may be eliminated,
effectively. Let us assume that
\begin{equation}
\Box \Lambda=-{\cal V}({\cal M})\Lambda
\end{equation}
for a spinless particle, where ${\cal V}$ is some function of mass.
If one assumes that  ${\cal M}$ is approaximately constant, then the term $e^{-\beta^2\Box}\Lambda$
can be written as
\begin{equation}
e^{-\beta^2\Box} \Lambda \simeq e^{\beta^2{\cal V}}\Lambda
\end{equation}
So the equation (29) reads
\begin{equation}
{\cal M}^2=m^2+\frac{\hbar^2 \Omega_0{\cal I}}{c^2}e^{\beta^2{\cal V}}
\end{equation}
or
\begin{equation}
{\cal V}({\cal M})=\frac{1}{\beta^2}\ln \left [ \frac{c^2}{\hbar^2\Omega_0 {\cal I}}({\cal M}^2-m^2) \right ]
\end{equation}
That is to say, it is possible to define an approximate effective action
as follows
\begin{equation}
{\cal A}=\int d^4y \left \{ \frac{1}{2}\left [ \partial _{\mu}{\cal S}\partial ^{\mu}{\cal S}-{\cal M}^2c^2\right ] +\frac{1}{2}\partial _{\mu}\Lambda \partial ^{\mu}\Lambda -{\cal V}({\cal M})\Lambda ^2\right \}
\end{equation}
instead of (28). The motion predicted by (85) is approximately as one predicted
by (28). It has the advantage that it does not contain high degree derivatives.
\item In (Shojai 1996), we extended the nonrelativistic quantum force derived from direct-particle-interaction,
to many particle systems. This extension can be done easily in the case of relativistic
quantum force. 
Suppose we have $s$ kind of particles. An obvious generalization of (49) is 
\[ Q(x_1, \cdots , x_s)=\frac{\Omega_0}{\rho(x_1, \cdots , x_s)} \]
\begin{equation}
\times \int \prod _{i=1}^s d^4y_i\ \sqrt{\rho(x_1, \cdots , x_s)}\ {\cal G}(x_1 , \cdots , x_s ; y_1, \cdots , y_s)\ \sqrt{\rho(y_1, \cdots , y_s)}
\end{equation}
An appropriate generalization of the kernel is
\[ {\cal G}_{Spin=0}(x_1 , \cdots , x_s ; y_1, \cdots , y_s)= \]
\begin{equation}
\exp \left \{ \sum_{i=1}^s \beta ^2(x_i-y_i)^2+\sum_{i>j=1}^s \beta'^2(x_i-x_j)_{\mu}(y_i-y_j)^{\mu} \right \}
\end{equation}
\[ {\cal G}_{Spin=\frac{1}{2}}(x_1 , \cdots , x_s ; y_1, \cdots , y_s)= \exp \left \{ \sum_{i=1}^s \beta^2 \left [(x_i-y_i)_{\mu}1+ \epsilon_{\mu}\right ]\left [(x_i-y_i)^{\mu}1+ \epsilon^{\mu}\right ] \right . \]
\begin{equation}
\left . +\sum_{i>j=1}^s \beta'^2\left [(x_i-x_j)_{\mu}1+ \epsilon'_{\mu}\right ]\left [ (y_i-y_j)^{\mu}1+ \epsilon'^{\mu}\right ] \right \}
\end{equation}
For higher spins, the kernel ${\cal G}$ can be written immediately either
similar to (87) (for integer spins) or similar to (88) (for half-integer spins).
\par
A very interesting result is that the spin--statistics relation would emerge
from the foregoing relativistic quantum potential. In the Copenhagen quantum mechanics,
the spin--statistics relation states that the wave-function of two identical bosons or fermions
is given by
\begin{equation}
\psi(x_1, x_2)=\frac{1}{\sqrt{2}}\left [ \psi_1(x_1)\psi_2(x_2)\pm\psi_1(x_2)\psi_2(x_1)\right ]
\end{equation}
where $+$ sign refers to bosons and $-$ sign refers to fermions, and $\psi_i(x_i)$
is the one particle wave-function of the $i$th particle. If one
chooses $x_1\simeq x_2=x$, one has
\begin{equation}
|\psi(x_1, x_2)|_{x_1\simeq x_2=x}\simeq \left \{ \begin{array}{lr}
|\psi_1(x)||\psi_2(x)| & for\ bosons,\\ 0 & for\ fermions. \end{array} \right .
\end{equation}
i.e. two bosons can be at the same point, but two fermions cannot. In the
de-Broglie--Bohm quantum theory of motion, the quantum force is responsible for
this behaviour. But in that theory, the symmetrization or antisymmetrization of the 
wave-function is made by hand.
In the direct-particle-interaction theory of the quantum force, this problem has a
logical answer.
For spin zero case, setting $x_1\simeq x_2=x$, the kernel does not couple $y_1$ and $y_2$, so if
$\Lambda (y_1, y_2)$ is decomposable as $\Lambda _1(y_1)\Lambda _2(y_2)$, then
the quantum potential too has this property. Thus, this decomposed state is stationary.
That is for bosons, the first part of (90) is applicable. For spin one-half case,
setting $x_1\simeq x_2=x$ does not lead to a decoupled kernel in terms of
$y_1$ and $y_2$. Thus the decomposed form of $\sqrt{\rho}$ is not stationary, i.e.,
if one assumes that at some time, $\sqrt{\rho}$ is decomposable, the quantum force, 
forces it to an undecmposable one. The only stationary solution is the second part
of (90). Therefore {\it the spin--statistics relation is a logical conclusion of the 
direct-particle-interaction theory of the quantum force\/}.
\end{enumerate}
\section{CONCLUSION}
\par
It is argued that, if one wants to remain faithful to the logic, the experiments
done in the beginings of this century lead him to the conclusion that there is some
new force -- the quantum force. Also it is shown that this new force is a result
of direct-particle-interaction of the particles of an ensemble (or in other words,
the interaction of particle with all of its possibilities.) The appropriate
direct-particle-interaction can be written for any spin. The theory has the following
properties:
\begin{enumerate}
\item It presents a relativistic causal theory for quantal behaviours.
\item It has no superluminal motion.
\item The two essential assumptions of the de-Broglie--Bohm quantum theory
of motion are derived in this theory.
\item If one restricts himself to the {\it observed quantities\/} of the Copenhagen quantum mechanics,
the presented theory is equivalent to the relativistic quantum mechanics, up to some corrections. But it is
not a new interpretation.
\item It presents a natural framework for understanding the problem of creation and annihilation of particles. 
\item It leads to the spin--statistics relationship.
\end{enumerate}
\vspace{0.5cm}
{\large APPENDIX A}\\
{\bf A.1 Nonrelativistic quantum force}
\par
In ref. (Shojai 1996), we have shown that the nonrelativistic quantum potential, can be
derived from the direct-particle-interaction given by (18). Equation (18) can be
simplified by using the relations
$$ \int d^3y\ \exp\left \{-\left [\vec{y}+\zeta(\vec{x}-\vec{a}_k(t))\right ]^2/\alpha _s^2 +y^2/\gamma ^2 -y^2/\gamma ^2\right \} =(\pi \alpha _s^2)^{3/2} \eqno(A.1.A)$$
$$\left (\frac{2}{\pi \alpha _s^2}\right )^{3/4} \exp \left [ -(\vec{x}-\vec{a}_k(t))^2/\alpha ^2_s\right ] \simeq \sqrt{\delta(\vec{x}-\vec{a}_k(t))} \eqno(A.1.B)$$
$$\sum ^N_{k=1}\sqrt{\delta(\vec{x}-\vec{a}_k(t))}=\sqrt{\sum_{k=1}^N\delta(\vec{x}-\vec{a}_k(t))} \eqno(A.1.C)$$
with choices
$$ \zeta=\frac{1}{2}\left [1+\sqrt{1-4\frac{\alpha _s^2}{\alpha _{\ell}^2}}\right ] \eqno(A.1.D)$$
and
$$ \gamma=\frac{\sqrt{2}\alpha _s}{\left [1-\sqrt{1-4\frac{\alpha _s^2}{\alpha _{\ell}^2}}\right ]^{1/2}} \eqno(A.1.E)$$
as
$$Q(\vec{x},t)\simeq U_0 \left [2 \left (1+\sqrt{1-4\frac{\alpha _s^2}{\alpha _{\ell}^2}}\right )\right ]^{-3/4}\int d^3y\ \sqrt{\frac{\rho(\vec{x}+\vec{y},t)}{\rho(\vec{x},t)}}\ \exp \left [-y^2 \frac{1-\sqrt{1
-4\frac{\alpha _s^2}{\alpha _{\ell}^2}}}{2\alpha_s^2} \right ]\eqno(A.1.F)$$
or as equations (19) and (20), where
$$\Omega _0=U_0\left [2 \left (1+\sqrt{1-4\frac{\alpha _s^2}{\alpha _{\ell}^2}}\right )\right ]^{-3/4}\eqno(A.1.G)$$
$$\beta^2=\frac{1-\sqrt{1-4\frac{\alpha _s^2}{\alpha _{\ell}^2}}}{2\alpha_s^2} \eqno(A.1.H)$$
The integral in (19) can be easily done by using the Gaussian representation
of Dirac's delta function. Then, one can use the Backer-Hausdorf lemma to convert
the result of the integration into equation (21). This is done in (Shojai 1996), but
here we represent another method which leads to the same result. It rests on the
smoothening of the density functions. Assuming that the scale of change of the 
density is larger than $1/\beta$ and 
expanding $\sqrt{\rho(\vec{y},t)}$
around $\vec{y}=\vec{x}$
$$\sqrt{\rho(\vec{y},t)}=\sum_{n=0}^{\infty}\sum_{r_1\cdots r_n=1}^3 \frac{1}{n!} (y-x)_{r_1}\cdots (y-x)_{r_n} \partial _{r_1}\cdots \partial _{r_n}\sqrt{\rho(\vec{x},t)} \eqno(A.1.I)$$
one obtains
$$Q(\vec{x},t)=\frac{\Omega_0}{\sqrt{\rho(\vec{x},t)}}\int d^3y \sum_{n=0}^{\infty}e^{-\beta ^2|\vec{y}-\vec{x}|^2}\sum_{r_1\cdots r_n=1}^3 \frac{1}{n!}(y-x)_{r_1}\cdots (y-x)_{r_n} \partial _{r_1}\cdots \partial _{r_n}\sqrt{\rho(\vec{x},t)}$$
$$=\frac{\Omega_0}{\sqrt{\rho}}\sum_{n=0}^{\infty}\sum_{r_1\cdots r_n=1}^3\frac{1}{n!}\left (\int d^3z \ e^{-\beta^2z^2}z_{r_1}\cdots z_{r_n}\right )\partial _{r_1}\cdots \partial _{r_n}\sqrt{\rho} \eqno(A.1.J)$$
If one defines
$${\cal J}_n(\alpha)=\int d^3z\ z_{r_1}\cdots z_{r_n} e^{-\beta ^2z^2+\vec{\alpha}\cdot \vec{z}} \eqno(A.1.K)$$
one has
$${\cal J}_n(0)=\left [\frac{\partial}{\partial \alpha_{r_1}}\cdots \frac{\partial}{\partial \alpha_{r_n}} {\cal J}_0(\alpha)\right ]_{\alpha=0} \eqno(A.1.L)$$
${\cal J}_0(\alpha)$ can be easily calculated as
$${\cal J}_0(\alpha)=\int d^3z\ e^{-\beta ^2z^2+\vec{\alpha}\cdot \vec{z}}={\cal I}e^{\alpha ^2/4\beta ^2} \eqno(A.1.M)$$
where
$${\cal I}=\int d^3u\ e^{-\beta ^2 u^2}=\left ( \frac{\pi}{\beta ^2} \right )^{3/2} \eqno(A.1.N)$$
Now ${\cal J}_n(0)$ can be evaluated
$${\cal J}_{2n+1}(0)=0 \eqno(A.1.O)$$
$${\cal J}_0(0)={\cal I} \eqno(A.1.P)$$
$${\cal J}_2(0)=\frac{{\cal I}}{2\beta ^2}\delta _{r_1 r_2}\ \ \ \ and\ so\ on.\eqno(A.1.Q)$$
Using these relations one obtains (21).
It must be noted here that the method presented in (Shojai 1996) leads to (21) but the factor $1/4\beta^2$ in the exponential 
would differ. This is because in (Shojai 1996) we first integrate (A.1.F) 
(with $\rho$ looked as Gaussian representation of the delta functions) and then use the Backer-Hausdorf lemma,
and finally smooth the density function. On the other hand, here we first smooth the density function
and then integrate. The results of these two methods slightly differ as it is
stated in the above. It seems that the method presented here is more appropriate,
because we can extend it to the relativistic case. {\it So we assume that the quantum potential is given by (19) for nonrelativistic
particles and by (49) for relativistic particles, and that the density function in these relations is 
smoothen out with the scale of change of $\rho$ larger than $1/\beta$}.  \\
\\
{\bf A.2 Relativistic quantum force}
\par
For spin zero case, calculations are completely similar to the previous part.
$$Q(x)= \frac{\Omega_0}{\sqrt{\rho(x)}}\int\ d^4y \ e^{\beta^2(x-y)^2}\sqrt{\rho(y)}$$
$$=\frac{\Omega_0}{\sqrt{\rho}}\sum_{n=0}^{\infty}\sum_{r_1\cdots r_n=0}^3\frac{1}{n!}\left (\int d^4z\ e^{\beta^2z^2}z^{r_1}\cdots z^{r_n}\right )\partial _{r_1}\cdots \partial _{r_n}\sqrt{\rho} \eqno(A.2.A)$$
$${\cal J}_0(\alpha)=\int d^4z\ e^{\beta ^2z^2+\alpha\cdot z}={\cal I}e^{-\alpha ^2/4\beta ^2} \eqno(A.2.B)$$
$${\cal I}=\int d^4u\ e^{\beta ^2 u^2}\eqno(A.2.C)$$
$${\cal J}_n(0)=\left [\frac{\partial}{\partial \alpha_{r_1}}\cdots \frac{\partial}{\partial \alpha_{r_n}} {\cal J}_0(\alpha)\right ]_{\alpha=0} \eqno(A.2.D)$$
$${\cal J}_{2n+1}(0)=0 \eqno(A.2.E)$$
$${\cal J}_0(0)={\cal I} \eqno(A.2.F)$$
$${\cal J}_2(0)=-\frac{{\cal I}}{2\beta ^2}g_{r_1 r_2}\ \ \ \ and\ so\ on.\eqno(A.2.G)$$
The result is (27). Note that, ${\cal I}$ in this case is not well-defined, but it can be absorbed in $\Omega_0$.
\par
For spin one-half case, calculations change slightly
$$Q(x)= \frac{\Omega_0}{2\rho(x)}\left [ \int d^4y\ \overline{\Gamma}(x)\ e^{\beta^2[(x-y)_{\mu}1+\epsilon_{\mu}][(x-y)^{\mu}1+\epsilon^{\mu}]}\ \Gamma(y) + C.C. \right ]$$
$$=\frac{\Omega_0 \overline{\Gamma}(x)}{\rho}\left [\sum_{n=0}^{\infty}\sum_{r_1\cdots r_n=0}^3\frac{1}{n!}\left (\int d^4z\ e^{\beta^2[z_{\mu}1+\epsilon_{\mu}][z^{\mu}1+\epsilon^{\mu}]}\ z^{r_1}\cdots z^{r_n}\right )\partial _{r_1}
\cdots \partial _{r_n}\Gamma + C.C. \right ] \eqno(A.2.H)$$
$${\cal J}_0(\alpha)=\int d^4z\ e^{\beta^2[z_{\mu}1+\epsilon_{\mu}][z^{\mu}1+\epsilon^{\mu}]+\alpha \cdot z1}={\cal I}e^{-\alpha ^2/4\beta ^2-\alpha \cdot \epsilon} \eqno(A.2.I)$$
$${\cal J}_0(0)={\cal I} \eqno(A.2.J)$$
$${\cal J}_1(0)=-{\cal I}\epsilon_{r_1} \eqno(A.2.K)$$
$${\cal J}_2(0)=-{\cal I}\left [\frac{g_{r_1 r_2}}{2\beta ^2}-\epsilon_{r_1}\epsilon_{r_2}\right ]\ \ \ \ ,\ etc.\eqno(A.2.L)$$
The result is (36). Calculations for the spin-one and the spin-two cases are  very similar to the spin-zero case.\\
\newline
{\large APPENDIX B}
\par
In this appendix, we evaluate the energy-momentum tensor for a system, whose lagrangian
contains high degree derivatives of fields. Consider a lagrangian as
$${\cal L}={\cal L}(x,\phi ,\phi_{;\mu _1},\phi_{;\mu _1\mu _2}, \cdots )\eqno(B.A)$$
where $;\mu _i$ represents differentiation with respect to $x_{\mu _i}$.
The equations of motion for such a lagrangian is
$$\frac{\partial {\cal L}}{\partial \phi}-\left ( \frac{\partial {\cal L}}{\partial \phi_{;\mu_1}}\right )_{;\mu_1}+\left ( \frac{\partial {\cal L}}{\partial \phi_{;\mu_1 \mu_2}}\right )_{;\mu_1 \mu_2}-\cdots =0 \eqno(B.B)$$
\par
Suppose we make an infinitesimal transformation
$$x_{\mu}\longrightarrow x_{\mu}+\xi_{\mu}\eqno(B.C)$$
then the change in ${\cal L}$ is 
$$\delta {\cal L}=\left (\frac{\partial {\cal L}}{\partial \phi}\right )\delta \phi + \left (\frac{\partial {\cal L}}{\partial \phi_{;\mu _1}}\right )\delta \phi_{;\mu_1} + \cdots 
+ \left (\frac{\partial {\cal L}}{\partial x^{\nu}}\right )\xi^{\nu} \eqno(B.D)$$
where by $\left (\frac{\partial {\cal L}}{\partial x^{\nu}}\right )$ we mean derivative of ${\cal L}$ with respect to $x^{\nu}$ when all of the quantities
$\phi$, $\phi_{;\mu_1}$, $\cdots$ are fixed, i.e.
$$\left (\frac{\partial {\cal L}}{\partial x^{\nu}}\right )=\frac{\partial {\cal L}}{\partial x^{\nu}}-\frac{\partial {\cal L}}{\partial \phi} \phi_{;\nu}-\frac{\partial {\cal L}}{\partial \phi_{;\mu_1}} \phi_{;\nu \mu_1}-\cdots \eqno(B.E)$$
In order to calculate $\delta {\cal L}$, we need $\delta \phi_{;\mu_1}$, $\delta \phi_{;\mu_1\mu_2}$, etc.
$$\delta \phi_{;\mu_1}=\frac{\partial \phi '(x')}{\partial x'^{\mu_1}}- \frac{\partial \phi (x)}{\partial x^{\mu_1}}=\frac{\partial x^{\alpha}}{\partial x'^{\mu_1}}\frac{\partial}{\partial x^{\alpha}}(\phi (x)+\delta \phi)-\phi_{;\mu_1}$$
$$=(\delta^{\alpha}_{\mu_1}-\xi^{\alpha}_{;\mu_1})(\phi_{;\alpha}+(\delta \phi)_{;\alpha})-\phi_{;\mu_1}\eqno(B.F)$$
So
$$\delta \phi_{;\mu_1}=(\delta \phi)_{;\mu_1}-\phi_{;\alpha}\xi ^{\alpha}_{;\mu_1}\eqno(B.G)$$
In the same way,
$$\delta \phi_{;\mu_1 \mu_2}=(\delta \phi)_{;\mu_1 \mu_2}-\phi_{;\alpha}\xi ^{\alpha}_{;\mu_1 \mu_2}-\phi_{;\mu_1 \alpha}\xi^{\alpha}_{;\mu_2}-\phi_{;\mu_2 \alpha}\xi^{\alpha}_{;\mu_1}\eqno(B.H)$$
and so on.
\par
It can be easily shown that the invariance of lagrangian, i.e. $\delta {\cal L}=0$, leads
to the following conservation law
$$\partial _{\mu_1}{\cal J}^{\mu_1}=0 \eqno(B.I)$$
where
$${\cal J}^{\mu_1}={\cal L}\xi ^{\mu_1}+\frac{\partial {\cal L}}{\partial \phi_{;\mu_1}}G+\frac{\partial {\cal L}}{\partial \phi_{;\mu_1 \mu_2}}G_{;\mu_2}-\left (\frac{\partial {\cal L}}{\partial \phi_{;\mu_1 \mu_2}}\right )_{;\mu_2}G$$
$$+\frac{\partial {\cal L}}{\partial \phi_{;\mu_1 \mu_2 \mu_3}}G_{;\mu_2 \mu_3}+\left (\frac{\partial {\cal L}}{\partial \phi_{;\mu_1 \mu_2 \mu_3}}\right )_{;\mu_3}G_{;\mu_2}-\left (\frac{\partial {\cal L}}{\partial \phi_{;\mu_1 \mu_2 \mu_3}}
\right )_{;\mu_2 \mu_3}G + \cdots \eqno(B.J)$$
and
$$G=\delta \phi-\phi_{;\nu}\xi ^{\nu} \eqno(B.K)$$
\par
If one chooses
$$ \delta \phi=0\ \ \ \ \ \ \ and \ \ \ \ \ \ \xi^{\nu}=constant\eqno(B.L)$$
the conserved current is the energy-momentum tensor. For example, the energy-momentum tensor
for a relativistic spinless particle can be derived from the lagrangian (28)
and the following relations
$$G=-\Lambda_{;}^{\nu}\eqno(B.M)$$
$$\frac{\delta {\cal L}}{\delta \Lambda_{;\mu_1\cdots \mu_{2n+1}}}=0\eqno(B.N)$$
$$\frac{\delta {\cal L}}{\delta \Lambda_{;\mu_1\cdots \mu_{2n}}}=\frac{1}{2}\Omega_0{\cal I}\Lambda \frac{1}{n!}\left ( \frac{1}{4\beta^2}\right )^n \left \{ g^{\mu_1\mu_2}\cdots g^{\mu_{2n-1}\mu_{2n}}+\ permutations \right \} \eqno(B.O)$$
as
$${\cal T}^{\mu_1 \nu}={\cal L}g^{\mu_1 \nu}-\rho \partial ^{\mu_1}{\cal S}\partial ^{\nu}{\cal S}-\frac{1}{4}\frac{\Omega_0 {\cal I}}{(4\beta^2)^2}\Lambda \Lambda_{;}^{\mu_1 \nu}$$
$$+\frac{1}{4}\frac{\Omega_0 {\cal I}}{(4\beta^2)^2}\Lambda_{;}^{\mu_1} \Lambda_{;}^{\nu} +\cdots \eqno(B.P)$$
\newpage
{\bf REFERENCES}\\
\newline
---Bohm, D. 1952a, Phys. Rev., {\bf 85}, 166.
\newline
---Bohm, D. 1952b, Phys. Rev., {\bf 85}, 180.
\newline
---Bohm, D., and Hiley, J. 1993, {\it "The Undivided Universe"\/}, Routledge.
\newline
---Holland, P.R. 1993a, {\it "The Quantum Theory of Motion"\/}, Cambridge University Press.
\newline
---Holland, P.R. 1993b, Phys. Rep., {\bf 224}, No. 3, 95.
\newline
---Lam, M.M., and Dewdney, C. 1994a, Found. of Phys., {\bf 24}, No. 1, 3.
\newline
---Lam, M.M., and Dewdney, C. 1994b, Found. of Phys., {\bf 24}, No. 1, 29.
\newline
---Ohanian, H.C. 1976, {\it "Gravitation and Spacetime"\/}, W.W. Norton \& Company.
\newline
---Ribari\v c, M., and \v Su\v ster\v si\v c, L. 1990, {\it Conservation Laws and Open Questions of Classical Electrodynamics"\/}, World Scientific Publishing Co. Pte. Ltd..
\newline
---Shojai, A., and Golshani, M. 1996, Direct-Particle-Interaction as the Origin of
the Quantal Behaviours, IPM-96-124, submitted for publication.
\end{bf}
\end{document}